# Circuit Theory for SPICE of Spintronic Integrated Circuits

Sasikanth Manipatruni, Dmitri E. Nikonov, *Senior Member IEEE* and Ian A. Young, *Fellow, IEEE*

*Abstract*— We present a theoretical and a numerical formalism for analysis and design of spintronic integrated circuits (SPINICs). The formalism encompasses a generalized circuit theory for spintronic integrated circuits based on nanomagnetic dynamics and spin transport. We propose an extension to the Modified Nodal Analysis technique for the analysis of spin circuits based on the recently developed spin conduction matrices. We demonstrate the applicability of the framework using an example spin logic circuit described using spin Netlists.

## I. INTRODUCTION

SPINTRONICS, the technology of control and manipulation of the spin state of electrons and nanomagnets, is one of the most promising approaches for beyond CMOS logic, memory and analog applications [1-5]. Several spin based devices have been proposed [6-19] with the possibility of logic-non-volatility, intrinsic directionality, higher logical efficiency (large fan-in/fan-out) and re-configurability. Combined with novel approaches for memory hierarchy [20-21] and logic architecture [21-23], spintronics may enable high performance, normally-off (with zero standby power) and instantly-on computing engines.

In the past few years there has been tremendous progress in spintronic devices and integration [2, 24-30] propelled by the advances in materials and fabrication techniques. In particular, the advances towards in plane, three terminal and majority gate spin transfer torque devices [31-34] have opened the possibility of spin logic devices which enable computation to be performed entirely in the magnetic and spin states of materials. While the proposed devices show promising trends for non-volatile operation, low energy-delay products and better logical efficiency, the suitability of the devices as components for large scale integration remains to be shown. In particular, significant advances in spin logic device, circuit and system design are still required in-order to fully understand the suitability of spin devices for general purpose computing.

The goal of this paper is to outline the principles for analysis of integrated spintronic circuits so that the physics of spin transport can be utilized by SPICE developers and subsequently by circuit and system designers for the exploration of spintronics for 'beyond CMOS computing'. We base the present work on the physical principles for spin transport developed over the past few years, first in collinear magneto-electronics for spin valves [35] and then non-collinear magneto-electronics [36-38] for spin transfer torque devices. In particular, we use the formalism for 4X4 spin conduction matrices, introduced in [18, 19], which enable the combined analysis of nanomagnets connected by spin transport channels. These were utilized for modeling of spin logic devices in [19]. We describe the circuit level transport models for spintronic devices from the physics of the spin transport through nanomagnets and nano-channels. We generalize the Kirchoff's conservation laws for spin circuits to include spin dissipation [36]. We then extend the principles of modified nodal analysis (MNA) [62] to spin circuits enabling Netlist based SPICE simulations.

## II. CONCEPTS OF VECTOR SPIN CURRENT, SPIN-VOLTAGE AND SPIN CONDUCTION MATRICES

To develop the formalism for spin conduction, let us consider a branch of generic circuit consisting of two nodes $N_1$, $N_2$ connected by a conduction element $B_{12}$ (Fig. 1 a, b).

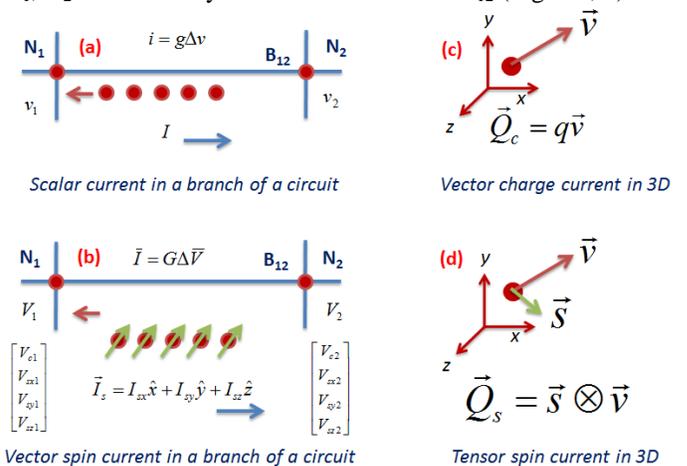

Fig. 1. Conceptual diagram of two nodes in a circuit connected by a conductance branch: a) two nodes connected by a scalar conductance in a regular circuit; b) two nodes connected by a spin conductance in a spin circuit. c) Conceptual diagram of a spin current tensor when a spin current flows in a 3D space. d) Spin current tensor is reduced to a spin current vector when a direction is implied by a branch of the circuit. The current and the voltages in a spin circuit are 4 component vectors carrying both the scalar current/voltage quantities and vector spin current/voltage quantities. The linearity of the circuit implies that the connecting branch is described by a 4X4 spin conductance matrix.

We now describe the concept of vector spin current $\vec{I}_s$ (with the three Cartesian components specified by three scalar numbers), vector spin voltage $\vec{V}_s$ (with three Cartesian components specified by three scalar numbers), total current vector $\bar{I}$ consisting of the coulomb current and vector spin current, total spin voltage $\bar{V}$ (consisting of the coulomb voltage and vector spin voltage) [18].

### A. Node of a spin circuit

We formally define the node of a spin circuit as a collection of physical points in a device or a circuit where all the quantities of interest for spin and charge transport are at equilibrium [36]. The mechanisms driving the node to the state of equilibrium are assumed to be much faster than the dynamics of the circuit.

### B. Vector Spin Current

Vector spin current in a branch of a spin circuit, is the net vector flow of magnetic moment along the branch of the circuit. It has the units of amperes. In general, the spin current flowing in a three dimensional space is a tensor [37, 39]. The spin tensor is described by a direction of the flow of the charges constituting the spin current and the direction of the net magnetic moment (spin) of the charges along each axis of the Cartesian coordinates (Fig. 1 c,d). However, in a circuit, the direction of the flow of charges is defined by the connectivity of the branch and therefore, the spin current flowing between two points of a spin circuit is a vector.

$$\vec{I}_s = I_{sx}\hat{x} + I_{sy}\hat{y} + I_{sz}\hat{z} \quad (1)$$

The spin current can be also related to the velocity and spin states of the carriers in a circuit/device. The components of the vector spin current are expressed as a sum over the momentum $k$ states of electrons, normalized to density of electrons,

$$I_{sj} = -e\sum_{A,k} Tr(\rho\hat{\sigma}_j)v_n \quad (2.1)$$

where $\sigma_j$ are Pauli matrices, $A$ is cross sectional area, $v_n$ is the velocity component normal to it, and $\rho$ is the spin density matrix. It is defined with the negative sign to reflect the negative charge of electrons, similarly to the electric charge current

$$I_c = -e\sum_{A,k} v_n \quad (2.2)$$

This way, the spin current corresponds to the flux of magnetic moments. Thus in Fig. 1a, the electrical current (blue arrow to the right) is opposite to the flux of electrons (red arrow to the left. If the net spin projection is positive, then the spin current, blue arrow in Fig. 1b, is opposite to the flux of electrons (red arrow).

### C. Vector Spin Voltage

Vector spin voltage at a node

$$\vec{V}_s = V_{sx}\hat{x} + V_{sy}\hat{y} + V_{sz}\hat{z} \quad (3)$$

can be understood intuitively as the state variable associated with the accumulation of spins of a certain direction. It is related to the *half*-difference in the electrochemical potentials of the electrons with their spin up and down along the direction of this vector. On the other hand, this difference can be related to the *half*-difference of density of the electrons with spin up and spin down, $\Delta n$, following the Valet-Fert theory [1, 35]:

$$-e\vec{V}_s = \Delta\vec{\mu} = (\Delta\mu_x\hat{x} + \Delta\mu_y\hat{y} + \Delta\mu_z\hat{z}) = \frac{4}{3}\frac{\Delta\vec{n}}{n}\mu \quad (4)$$

Where $n$ is the total density, $\vec{\mu}$ is the electrochemical potential. Spin voltage is defined with the negative sign as well. It is done to make the definition consistent to the usual electrochemical potential for electrons, which includes the term of voltage with a negative sign. Overall, this choice of signs in the definitions of spin current and spin voltage makes most of the relationships similar to those between charge current and voltage as in Fig. 2.

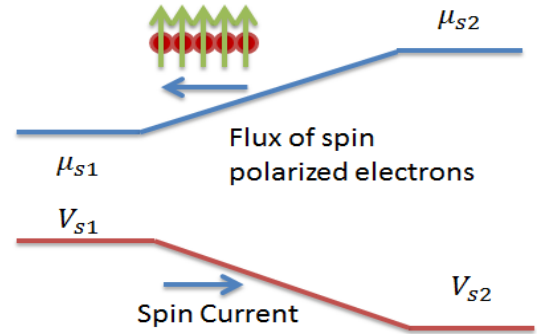

Fig. 2. Accumulation of spin up electrons is expressed in a higher spin electrochemical potential. Diffusion current of up spin electrons flows from a higher to a lower spin electrochemical potential. Equivalently, spin current flows from higher spin voltages to lower spin voltages.

### D. Total Spin Current Vector

The total spin current is simply the combination of the charge current and vector spin current. It is a 4x1 column vector.

$$\bar{I} = \begin{bmatrix} I_c & I_{sx} & I_{sy} & I_{sz} \end{bmatrix}^T \quad (5)$$

The ratio between the charge current and the magnitude of the spin current is the current's spin polarization ratio

$$\alpha_c = \frac{|\vec{I}_s|}{I_c} \quad (6)$$

### E. Total Spin Voltage Vector

The total spin voltage vector ($V$) is the combination of the scalar columbic potential and the vector spin potential. It is a 4x1 column vector.

$$\bar{V} = \begin{bmatrix} V_c & V_{sx} & V_{sy} & V_{sz} \end{bmatrix}^T \quad (7)$$

The ratio between the scalar potential and the magnitude of the spin potential is the voltage-spin polarization ratio of a node.

$$\beta = \left|\frac{e\vec{V}_s}{\mu}\right| = \frac{4}{3}\frac{|\Delta\vec{n}|}{n} \quad (7)$$

*F. Spin conduction matrix*

The concept of spin conduction matrix can be derived by postulating the linear response of current to voltage. This assumption is expected to be valid in metals, where electron density is high and electric fields change slowly compared to the scattering time. In the worst case, one has to take the branches of the circuit small enough to make it a good approximation. The spin Ohm's law, the linear relationship of spin voltage to spin current is

$$\bar{I} = G\Delta\bar{V} \qquad (8)$$

Where G is the 4X4 conductance matrix:

$$G = \begin{bmatrix} G_{11} & G_{12} & G_{13} & G_{14} \\ G_{21} & G_{22} & G_{23} & G_{24} \\ G_{31} & G_{32} & G_{33} & G_{34} \\ G_{41} & G_{42} & G_{43} & G_{44} \end{bmatrix} \qquad (9)$$

G can also be interpreted as a 4X4 tensor that scales and reorients the voltage vectors to obtain the spin current vector. Therefore, the spin conductance matrix of a conductance element is the matrix proportionality constant relating the vector spin current through an element with the vector spin voltage difference applied across a conductance element. In general the 16 components of a spin conduction matrix are non-zero and are set by the magnetic and geometric properties of the spin conductance element.

| Table 1. List of variables for spin circuit theory | | | |
|---|---|---|---|
| Variable | Notation | Matrix Size | Units (SI) |
| Vector Spin Current | $\vec{I}$ | 3X1 | A |
| Total Vector Spin Current | $\bar{I}$ | 4X1 | A |
| Vector Spin voltage | $\vec{V}$ | 3X1 | V |
| Total Vector Spin voltage | $\bar{V}$ | 4X1 | V |
| Spin Conductance | G | 4X4 | Ohm$^{-1}$ |
| Spin Population Polarization | $\Delta n$ | 3X1 | m$^{-3}$ |
| Spin Current Polarization | $\alpha_c$ | 1X1 | Unit less |
| Spin Voltage Polarization | $\beta$ | 1X1 | Unit less |

### III. CONSERVATION LAWS FOR SPIN CIRCUITS

We now describe the extension of the Kirchoff's current and voltage laws to spin circuits [36]. Establishing the conservation laws is essential to setup a unique set of equations governing the currents and voltages.

*A. Kirchhoff's voltage law extension for spin circuits*

The traditional voltage law for circuits is extended straightforwardly to spin circuits: since the sum of voltage differences in any closed loop is zero:

$$\sum_{i,j \in S_{loop}} \bar{V}_i - \bar{V}_j = 0 \qquad (10)$$

where $S_{loop}$ is the set of all node pairs in a given closed loop.

*B. Kirchoff's current law extension for spin circuits*

Apart from the vectorial nature of spin currents and voltages, spin circuits from electric circuits is that charge is strictly conserved, but spin is not. We handle the non-conservative nature of the spin currents entering a node by introducing a spin dissipation current to a virtual ground [19]. The traditional Kirchhoff's current law is thus extended to spin circuits as follows: the sum of the vector spin currents entering node is equal to the total dissipated vector spin current at the node. At node *i*, the spin node current law is given by equation 11, where $\vec{I}_{ij}$ is the spin current from node *i* to node *j*; B$_i$ is the set of all nodes connected to node *i*; $\bar{I}_{d,i}$ is the total spin current dissipated due to spin flip events happening at the node ($I_{sf}$).

$$\sum_{j \in B_i} \vec{I}_{ij} - \vec{I}_{d,i} = 0 \qquad (11)$$

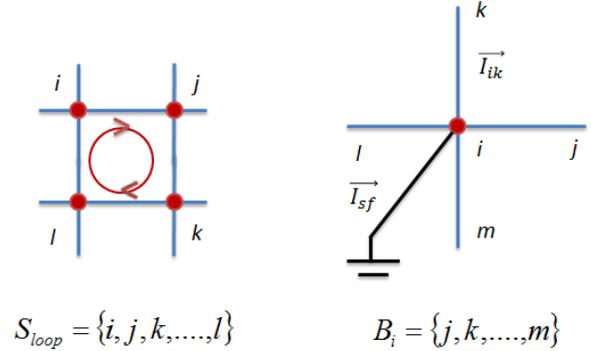

$S_{loop} = \{i, j, k, ...., l\}$          $B_i = \{j, k, ...., m\}$

Fig. 3. Conservation laws for spin voltages and currents: a) The sum of loop voltage differences is zero. b) The sum of the physical spin currents from all physical branches is equal to the spin flip current to a virtual ground.

### IV. 4-COMPONENT SPIN CONDUCTION MATRICES FOR NON-MAGNETIC CHANNELS, SERIES AND PARALLEL ELEMENTS

We describe spin conduction matrices for non-magnetic elements (NME).

*A. Spin conduction through non-magnetic elements*

The spin conduction through a linear non-magnetic conductive element can be described as follows: a) the charge current through the device is directly proportional to the applied scalar voltage difference b) the spin vector current though the element is directly proportional and collinear to the vector spin voltage difference applied to the NME, i.e.

$$I_c = g\Delta V_c \qquad (12)$$

$$\vec{I}_s = g_s \Delta \vec{V}_s \qquad (13)$$

where g$_s$ is a scalar quantity, $\Delta\vec{V_s}$ is the applied spin vector voltage across the NME, g is the scalar conductance, $\Delta V_c$ is the applied voltage difference.

*B. Spin conduction matrix for a spin elastic series branch*

Using the concept of spin conduction through non-magnetic elements, we write the conductance of a series resistor as [18]

$$G_{se} = \begin{bmatrix} g & 0 & 0 & 0 \\ 0 & g_s & 0 & 0 \\ 0 & 0 & g_s & 0 \\ 0 & 0 & 0 & g_s \end{bmatrix} \quad (14)$$

where we described the matrix elements earlier. A resistor with no spin flip is described fully by a single element as shown in Eq. (14). However, a series resistor with spin-flip needs to include a spin flip conductance to accommodate for the loss of spin current.

### C. Spin conduction matrix for a spin-flip conductance

We use the concept of spin-flip conductance to handle a non-spin-ballistic resistor. The conductance of a spin flip resistor is [18]

$$G_{sf} = \begin{bmatrix} 0 & 0 & 0 & 0 \\ 0 & g_{sh} & 0 & 0 \\ 0 & 0 & g_{sh} & 0 \\ 0 & 0 & 0 & g_{sh} \end{bmatrix} \quad (15)$$

where $g_{sh}$ is the spin flip conductance to accommodate the loss of spin polarization. The spin flip conductance sinks the spin current to a virtual spin ground to emulate spin current conservation even in presence of spin flip events.

### D. Spin conduction model (Π- or T- distributed equivalent) for distributed channel with spin flip

The spin conduction model for distributed channels in Π-equivalent and T-equivalent networks can be expressed using the shunt and series conductances described above. Let us consider a non-magnetic channel connecting two nodes of a spin circuit $N_1$ and $N_2$, Fig. 4(a). The process of spin flip which causes a loss of spin current from the channel is modeled using shunt resistances which go to the spin voltage ground [0 0 0 0]$^T$, Fig. 4(b). No charge current flows through the shunt elements since the conductance components $G_{1i}$ are zero.

The equivalent conductances in the Π-equivalent circuit for a non-ballistic elastic channel with cross section area of the channel $A_{ch}$, channel resistivity $\rho_{ch}$, the channel length $L_c$, spin-flip length of the channel material $\lambda_s$, is shown in Appendix A.

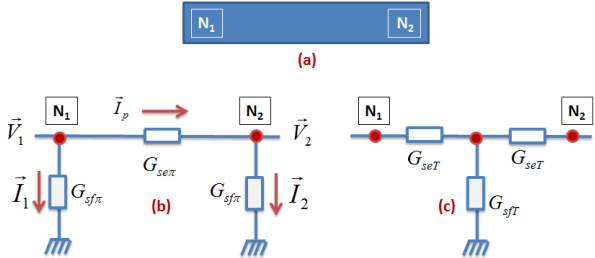

Fig. 4. Π and T Equivalent circuits for a normal metal distributed channel. a) Normal metal channel connecting nodes $N_1$ and $N_2$ b) Π equivalent circuit representing the distributed channel c) T equivalent circuit representing the distributed channel.

Depending on the ground conditions, a T-model (Fig. 4(c)) may be more convenient. The conversion from Π-equivalent to a T-equivalent is as follows (see Appendix B for details):

$$G_{seT} = G_{sf\pi} + 2G_{se\pi} \quad (16)$$

$$G_{sfT} = 2G_{sf\pi} + G_{sf\pi}G_{se\pi}^{-1}G_{sf\pi} \quad (17)$$

## V. 4-COMPONENT SPIN CONDUCTION MATRICES FOR MAGNETIC ELEMENTS

### A. Spin conduction through magnetic elements

In contrast to spin conduction through non-magnetic elements, spin conduction through magnetic elements in general can have; (a) coupling between scalar voltages and spin currents and (b) the spin current can be non-collinear to the vector spin voltage difference.

We next describe the conduction from a ferromagnet to a normal metal.

### B. Spin conduction at ferromagnet and normal metal interface

Spin conduction from a ferromagnet to a normal metal can be understood as the spin dependent current in response to spin voltages. The spin voltage at a node is in turn a result of a spin polarized population set up via spin injection from elsewhere [35]. Microscopically, the conduction is happening via spin dependent reflection and transmission at the interface of the ferromagnet and the normal metal [40-41]. Much of the formalism is derived from quantum transport scattering theory, see e.g. [42], and draws on the work from superconductive transport [43]. We refer the reader to [37] for a detailed physical explanation.

We first derive the 4X4 conduction matrix of a ferromagnet (FM) to normal metal (NM) interface (Fig. 5) from the spin conduction equations [38]. The elements of the spin conduction matrix can be filled with experimental properties. Let the 4X1 spin voltages at the FM and NM be $\bar{V}_{FM} = [V_F; 0]$ and $\bar{V}_{NM} = [V_N; \vec{V}_{sp}]$ respectively. Here for simplicity we ignore the spin accumulation in FM. Let $\bar{I}_{FN} = [I_c; \vec{I}_s]$ be the 4X1 spin current from FM to NM and $\hat{m}$ be the vector direction of the magnet's magnetic moment. Then, according to [38], the charge current is

$$I_C = G(V_N - V_F) + \alpha G \hat{m} \bullet \vec{V}_{sp} \quad (18)$$

The total spin current is given by:

$$\vec{I}_s = \vec{I}_\parallel + \vec{I}_\perp \quad (19)$$

$$\vec{I}_\parallel = G\left(\alpha(V_N - V_F) + \hat{m} \bullet \vec{V}_{sp}\right)\hat{m} \quad (20)$$

$$\vec{I}_\perp = G_{SL}\left[\hat{m} \times \left[\vec{V}_{sp} \times \hat{m}\right]\right] + G_{FL}(\vec{V}_{sp} \times \hat{m}) \quad (21)$$

where $\vec{I}_\parallel$ is the component of the spin current parallel to the magnetic moment and $\vec{I}_\perp$ is the spin current perpendicular to the magnetic moment. The expressions for spin torque conductances $G_{SL}$, and $G_{FL}$, which are related to the spin reflection and transmission properties of the interface, are described in Appendix C.

The spin conduction matrix elements can also be deduced from experimental properties of the FM-NM interface. G is the value of the total conductance of the interface, α is the spin selectivity of the FM-NM interface. $G_{SL}$ is the spin transfer conductance of the normal metal. The Sharvin conductance is the quantum limit of this term. $G_{FL}$ is the field-like conductance term which is typically zero for many metal interfaces. It is encountered at higher voltages in tunneling barriers adjacent to ferromagnets.

### C. Spin conduction matrix of a fixed nanomagnet

We derive the conduction matrix in a special case of the magnetization parallel to the plane of the interface, as shown in Fig. 5. We choose the coordinate system such that the x-axis is along the direction of magnetization $\hat{m} = \hat{x}$ and the other two axes form a right-handed coordinates. Then using projections to these coordinates, equations (18) - (21) become;

$$I_C = G_{11}(V_N - V_F) + \alpha G_{11} V_{sx} \quad (22)$$

$$I_{sx} = \alpha G_{11}(V_N - V_F) + G_{11} V_{sx} \quad (23)$$

$$I_{s\perp} = G_{SL}(V_{sy}\hat{y} + V_{sz}\hat{z}) + G_{FL}(V_{sz}\hat{y} - V_{sy}\hat{z}) \quad (24)$$

Hence, the generalized Ohm's law for the FM-NM interface is [18];

$$\begin{bmatrix} I_c \\ I_{sx} \\ I_{sy} \\ I_{sz} \end{bmatrix} = \begin{bmatrix} G_{11} & \alpha G_{11} & 0 & 0 \\ \alpha G_{11} & G_{11} & 0 & 0 \\ 0 & 0 & G_{SL} & G_{FL} \\ 0 & 0 & -G_{FL} & G_{SL} \end{bmatrix} \begin{bmatrix} V_N - V_F \\ V_{sx} \\ V_{sy} \\ V_{sz} \end{bmatrix} \quad (25)$$

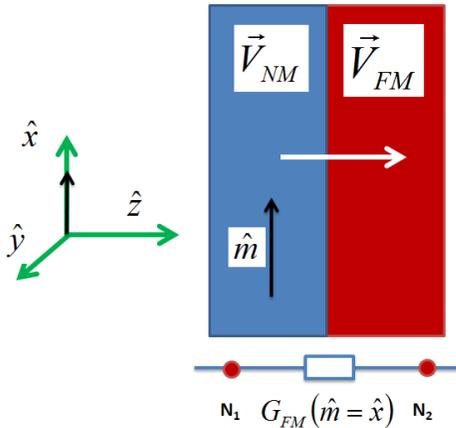

Fig. 5. Circuit model for spin transport between a ferromagnet (FM) and a normal metal (NM). The entire FM is treated as a node with a specific vector spin voltage.

$$G_{FN}(\hat{x}) = \begin{bmatrix} G_{11} & \alpha G_{11} & 0 & 0 \\ \alpha G_{11} & G_{11} & 0 & 0 \\ 0 & 0 & G_{SL} & G_{FL} \\ 0 & 0 & -G_{FL} & G_{SL} \end{bmatrix} \quad (26)$$

Since Eqs. (18)-(21) are independent of the orientation of the interface, we can extend by induction that the same expression for the conduction matrix is valid for any direction of magnetization provided that the coordinate system has its x-axis aligned to it, $\hat{m} = \hat{x}$:

### D. Spin conduction matrix of a free layer nanomagnet

Now we need the expression of the spin conduction matrix of the FM-NM interface in the fixed coordinate system tied to the nanomagnet shape (e.g elliptical, see Fig. 6) rather than the instantaneous direction of magnetization. This derivation is given in Appendix D. The 4X4 conduction matrix in the fixed coordinate system given by

$$G_{FN}(\hat{m}) = R^{-1}(\hat{m}) G_{FN}(\hat{x}) R(\hat{m}) \quad (27)$$

where R is the rotation matrix

$$R(\hat{m}) = \begin{bmatrix} 1 & 0 & 0 & 0 \\ 0 & r_{22} & r_{23} & r_{24} \\ 0 & r_{32} & r_{33} & r_{33} \\ 0 & r_{42} & r_{43} & r_{44} \end{bmatrix} \quad (28)$$

The elements of the rotation matrix R are defined by the expression of the unit vectors of the axes (X, Y, Z) tied to the magnetization in terms of the fixed coordinates (x,y,z)

$$[r_{22} \quad r_{23} \quad r_{24}] = \hat{X} = \hat{m} \quad (29)$$

$$[r_{32} \quad r_{33} \quad r_{34}] = \hat{Y} = -(\hat{X} \times \hat{x})/\|\hat{X} \times \hat{x}\| \quad (30)$$

$$[r_{42} \quad r_{43} \quad r_{44}] = \hat{Z} = \hat{X} \times \hat{Y} \quad (31)$$

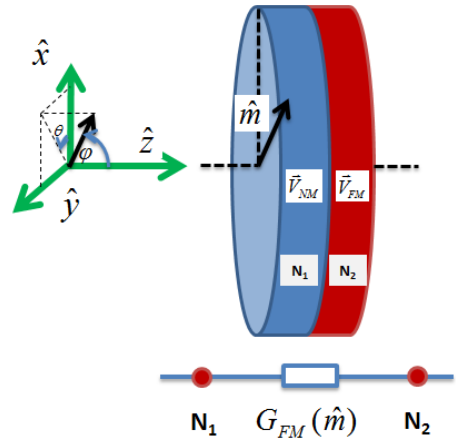

Fig. 6. Circuit model for spin transport between a ferromagnet (FM) and a normal metal (NM). The FM magnetization can point in any direction in three dimensions as determined by the nanomagnet dynamics

| Table 2. Transport parameters used in spin circuit theory | | | |
|---|---|---|---|
| Variable | Notation | Typical value | Units (SI) |
| Spin flip length of ferromagnetic metals | $\lambda_s$ | 5 (NiFe, Py) – 50 (Co) [31, 34, 44] | nm |
| Spin flip length of normal metals | $\lambda_n$ | 200-1000 [31, 34, 44-47] | nm |
| Spin current polarization | $\alpha_c$ | 0.3-0.5 [34, 44-47] | - |
| Sharvin resistance of a normal metal | $G_{sh}$ | 0.47 (Co)-0.58 (Cu) [48] | Ohm$^{-1}$ m$^{-2}$ |

## VI. SELF-CONSISTENCY OF NANOMAGNET DYNAMICS WITH SPIN CIRCUIT ANALYSIS

We now describe a coupled spin transport-magnetization dynamics model [76] for solving spin integrated circuits which employ nanomagnets for spin injection. The phenomenological equation describing the dynamics of nanomagnet with a magnetic moment unit vector ($\hat{m}$), the modified Landau-Lifshitz-Gilbert (LLG) equation [49-50], with spin transfer torques in the form of [38] is (see Table 3 for parameters)

$$\frac{\partial \hat{m}}{\partial t} = -\gamma \mu_0 [\hat{m} \times \vec{H}_{eff}] + \alpha \left[ \hat{m} \times \frac{\partial m}{\partial t} \right] + \frac{\vec{I}_\perp}{eN_s} \quad (32)$$

where γ is the electron gyromagnetic ratio; $\mu_0$ is the free space permeability; $\vec{H}_{eff}$ is the effective magnetic field due to material/geometric/surface anisotropy; α is the Gilbert damping of the material and $\vec{I}_\perp$ is the component of vector spin current perpendicular to the magnetization ($\hat{m}$) leaving the nanomagnet, $N_s$ is the total number of Bohr magnetons per magnet. $\vec{I}_\perp$ can also be rewritten as $\vec{I}_\perp = \vec{I}_s - \hat{m}(\hat{m}.\vec{I}_s) = \hat{m} \times (\vec{I}_s \times \hat{m})$. The positive sign of the spin torque term is related to the fact (Section II.B) that the spin current is has the same direction as the flux of magnetic moments. Implicit in the LLG equation is the fact that absolute values of the magnetic moments of single domain nanomagnets remain constant. The noise properties of nanomagnets play a critical role in the dynamics of the magnets [51-54]. See Appendix F for a description of the noise properties and numerical methods for stochastic LLG equations.

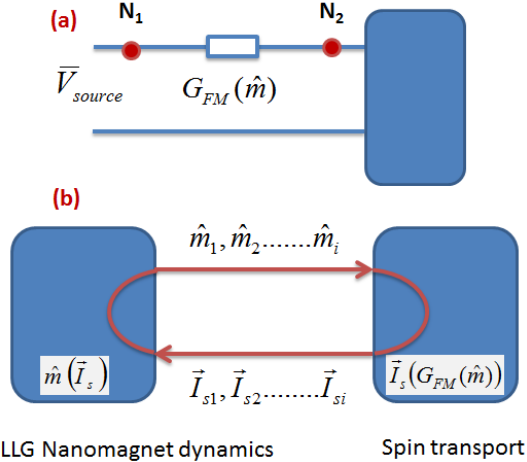

Fig. 7. Self-consistency of nanomagnet dynamics with spin transport a) example circuit demonstrating the need for self-consistency b) Self-consistency between LLG dynamics and spin transport.

In general, the direction of the nanomagnet magnetic moments of a spin circuit and the spin transport via a spin circuit are coupled together. The spin current entering a nanomagnet is defined by the conductance of the nanomagnet at the present angular position. This is because the equivalent conductance of the nanomagnet is determined by the direction the nanomagnet's moment. Consider an example spin circuit shown in Fig 7(a). The current passing through the circuit depends on the direction of the magnet ($\hat{m}$) while the direction of the magnet is modified depending on the injected spin current. Hence at each instant of time a self-consistent solution needs to be calculated to ensure accuracy. Fig. 7 b shows the self-consistent loop between LLG nanomagnet dynamics and spin transport [76]. The LLG solvers pass the condition of the magnets to the spin circuit and the spin circuit solver passed the spin vector current to the LLG solver at each pass of the self-consistent loop till a solution is reached. Self-consistency can also be addressed by using an implicit numerical solver [55].

*Table 3. Nanomagnet parameters for spin circuit theory used in combination with a macrospin nanomagnet model*

| Variable | Notation | Value/Typical Value | Units (SI) |
|---|---|---|---|
| Free Space Permeability | $\mu_0$ | $4\pi \times 10^{-7}$ | $JA^{-2}m^{-1}$ |
| Gyromagnetic ratio | γ | $17.6 \times 10^{10}$ | $s^{-1}T^{-1}$ |
| Saturation Magnetization of the Magnet | $M_s$ | $10^6$ [56] | A/m |
| Damping of the Magnet | α | 0.007-0.01 [57-59] | - |
| Barrier Height | $E_b$ | 40-100 [60] | kT |
| Effective Internal Anisotropic Field | $H_{eff}$ | $10^3 - 10^6$ [61] | A/m |
| Number of Bohr magnetons in the nanomagnet | $N_s$ | $10^3 - 10^6$ | - |

## VII. MODIFIED NODAL ANALYSIS FOR SPIN CIRCUITS

We now extend the modified nodal analysis (MNA) to spin circuits in order to provide a scalable way to analyze multi-node systems. A computational method for solving spin circuits is essential even for few node circuits (nodes >2) since the ordering of the spin conduction matrices is crucial while calculating equivalent conductances (see Appendix G). Following closely the formalism for MNA [62], the spin-MNA solves the following equation:

$$[X] = [A]^{-1}[Z] \quad (33)$$

where $A$ is a matrix formed based on the connectivity of the circuit, location of the voltage and current sources; $X$ is the vector comprising unknown node voltages and unknown currents through the voltage sources; $Z$ is the vector comprising of the voltages of the voltage sources and currents at the current sources. A can be also be written as

$$A = \begin{bmatrix} G & B \\ C & D \end{bmatrix} \quad (34)$$

where matrix $A$ has the size $4(m+n) \times 4(m+n)$ (n is the number of nodes, and m is the number of independent voltage sources). Matrix $G$ has the size $4n \times 4n$ and is determined by the interconnections between the passive circuit elements. Matrix $B$ has the size $4n \times 4m$ and is determined by the connection of the voltage sources. Matrix $C$ has the size $4m \times 4n$ and is determined by the connection of the voltage sources. ($B$ and $C$ are closely related, particularly when only independent sources are considered). Matrix D is $4m \times 4m$ and is zero if only independent sources are considered.

We note that the spin currents entering the magnets can be extracted as the Cartesian components of the total spin current. For a magnet connected between $i^{th}$ and $j^{th}$ nodes, the spin current entering $i^{th}$ node

$$\vec{I}_{ij} = G(i,j)(\vec{V}_i - \vec{V}_j) \quad (35)$$

For a description of the rules/algorithm for writing matrix $A$ please see Appendix H [63]. Matrix A can be assembled as shown in Fig. 8. The proposed method can handle a combination of nonmagnetic and magnetic elements as well as dependent and independent spin/regular voltage and current sources. For a detailed description on the regular MNA algorithm and dependent sources please see [64, 65]. The solution of the spin MNA equation can be simplified by optimal ordering of the equations to obtain sparsity [66] among other optimization techniques. For an algorithm for parsing a netlist to do MNA see for example [65].

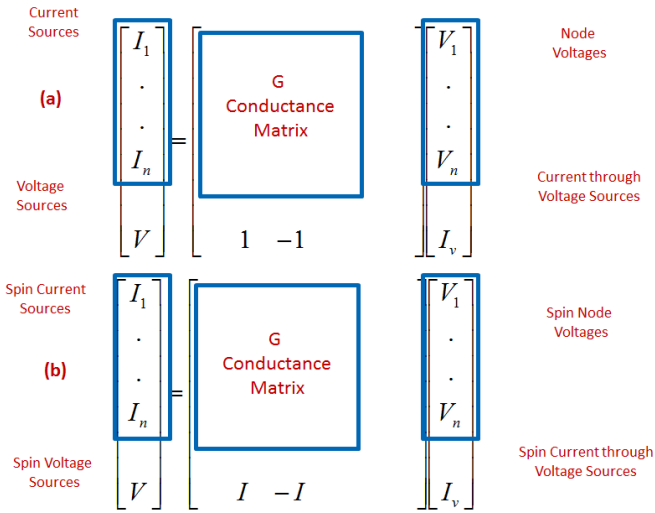

Fig. 8. a) Constitution of the MNA matrix equation for normal circuits b) Constitution of the Spin MNA matrix equation for spin circuits.

| Table 4. Spin MNA Matrices | | |
|---|---|---|
| Variable | Size in MNA | Size in Spin-MNA |
| X | $(m+n)X(m+n)$ | $(4m+4n)X(4m+4n)$ |
| A | $(n+m)X1$ | $(4n+4m)X1$ |
| Z | $(n+m)X1$ | $(4n+4m)X1$ |
| G | $nXn$ | $4nX4n$ |
| B | $nXm$ | $4nX4m$ |
| C | $mXn$ | $4mX4n$ |
| D | $mXm$ | $4mX4m$ |

## VIII. SIMULATION OF AN EXAMPLE SPIN CIRCUIT

We now describe an example spin circuit with embedded nanomagnets to demonstrate the effectiveness of the proposed framework. Let us consider a non-local spin injection detection device shown in Fig. 9 [19, 31, 34]. We consider a lateral spin injection-detection device which has been proposed recently as an all spin logic device. The device consists of two nanomagnets communicating via a non-magnetic channel. The device operates as an inverting gate for positive applied voltages and a non-inverting gate for negative applied voltages.

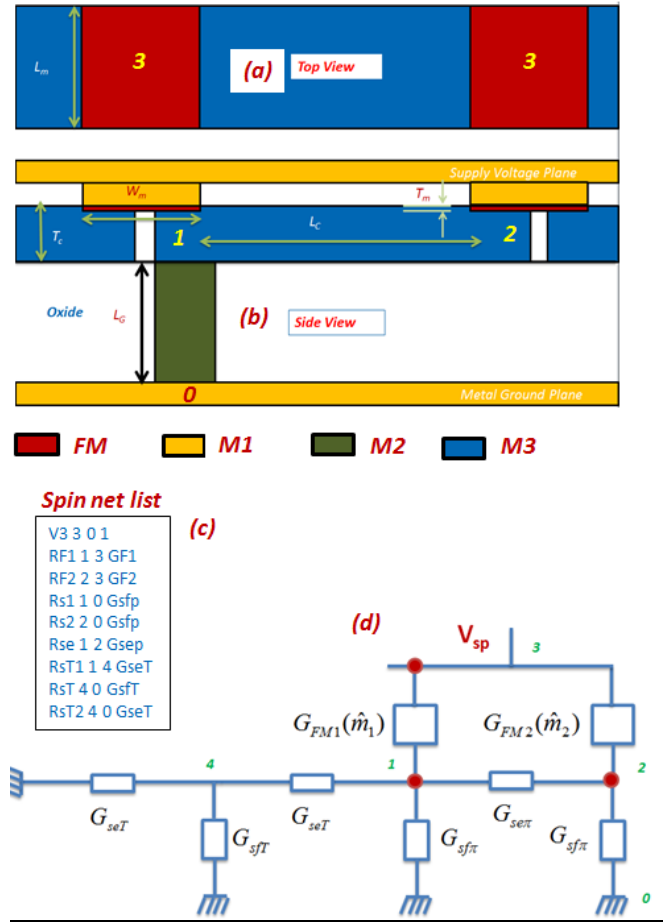

Fig. 9. A lateral spin logic device comprising of two nanomagents and non-magnetic channels. Channel connecting 1-2 acts as an interconnect between the two magnets transporting spin polarized currents. a) Top view of b) Side view of a lateral spin logic device c) Netlist of the circuit that can be parsed by a spin-MNA algorithm d) Circuit model of spin logic device.

| Table 5. Parameters used for example circuit simulation | | | |
|---|---|---|---|
| Variable | Notation | Value | Units (SI) |
| Saturation Magnetization of the Magnet | $M_s$ | $10^6$ | A/m |
| Damping of the Magnet | $\alpha$ | 0.007 | - |
| Effective Internal Anisotropic Field | $H_{eff}$ | $3.06 \times 10^4$ | A/m |
| Barrier of the magnet | $\Delta/kT$ | 40 | |
| Length of Magnet | $N_s$ | $10^3 - 10^6$ | - |
| Thickness of Magnet | $T_m$ | 3 | nm |
| Width of Magnet | $W_m$ | 37.8 | nm |
| Length of Magnet | $L_m$ | 75.7 | nm |
| Length of channel | $L_c$ | 100 | nm |
| Thickness of channel | $T_c$ | 200 | nm |
| Length of ground lead | $L_g$ | 200 | nm |
| Thickness of ground lead | $T_g$ | 100 | nm |
| Channel conductivity | $\rho$ | $7 \times 10^{-9}$ | $\Omega.m$ |
| Sharvin conductivity | $G_{sh}$ | $0.5 \times 10^{15}$ | $\Omega.m^{-2}$ |
| Polarization | $\alpha_c$ | 0.8 | |

Intuitively, the operation of the device can be explained as follows: the magnets create spin polarized population densities underneath the magnets and setup spin diffusion currents through the channel. The direction of this spin diffusion current is set by the relative strength of the spin polarization of the carriers. For a ground terminal set near the input magnet it can be shown that the magnet 1 acts as a fixed magnetic terminal, while the second magnet responds to the spin diffused to beneath it, depending on the applied voltages. For a positive applied voltage, the device shown in Fig. 9 acts like an inverting gate, where the output becomes a logical invert of the input. For negative applied voltages, the output becomes a copy of the input magnet's condition. The sectioned structure of the channel is required to isolate spin logic gates, where the interconnection between gates (concatenation) is achieved via a continuous free layer magnet. The non-reciprocity of (output to input signal transport) spin logic comes from an asymmetry between input and output magnets. This asymmetry can be achieved via a) an asymmetric overlap of the magnet such that the area of the output is greater than the area of input magnet b) asymmetric ground condition c) asymmetric spin injection efficiency (Fig 9) d) asymmetric spin damping constant

### A. Example spin circuit: Numbering the nodes, forming spin Netlist

We model the device as a spin circuit comprising of two nanomagnets and non-magnetic conductive elements. The non-magnetic elements model the behavior of the metal channels connecting the magnets to each other and to the ground. In Fig. 9, we show the top view and side view of the device. We choose the node-0 to be the ground and number the remaining nodes as per the convention of MNA. Node 1, 2 represent the points in the device just below the magnets representing the ends of the channel. Node 3 is common node shared by the magnets and the supply. We can now derive the circuit diagram for the device as shown in Fig. 9 (d). The magnetic elements are represented by $G_{FM1}$ & $G_{FM2}$; the non-magnetic channel is represented by a Π-equivalent circuit as described in section IV-D. The ground connection branch $B_{10}$ is represented by a T equivalent circuit. The assumed dimensions and the list of variables are shown in Table 5. We show the netlist for the device in Fig 9 c.

### B. Example spin circuit: Forming the MNA equation

Using the rules described in Appendix H, we can build the spin-MNA equation for the circuit in Fig. 9 d as follows: the G matrix is filled with the spin conductivity elements connecting to nodes 1-4. The elements representing connectivity A(5,3) is filled with an identity matrix such that the applied voltage at node 3 is $V_{sp}$. The row A3 represents the KCL at the node 3 and correspondingly has an identity matrix at element A(3,5).

$$\begin{bmatrix} O \\ O \\ O \\ O \\ V_{sp} \end{bmatrix} = \begin{bmatrix} G_{FM1}+G_{sh1}+G_{Se} & -G_{Se\pi} & -G_{FM1} & O & O \\ -G_{Se\pi} & G_{FM2}+G_{sf\pi}+G_{Se\pi} & -G_{FM2} & O & O \\ -G_{FM1} & -G_{FM2} & G_{FM1}+G_{FM2} & O & I \\ -G_{seT} & O & O & G_{seT}+G_{seT}+G_{sfTt} & O \\ O & O & I & O & O \end{bmatrix} \begin{bmatrix} V_1 \\ V_2 \\ V_3 \\ V_4 \\ I_{sp} \end{bmatrix}$$ (36)

### C. Example spin circuit: Self consistent solution of spin-MNA with LLG equations

The self-consistent solution is obtained by solving all the nanomagnet dynamical equations consistently with the transport equations (Fig 8). The full set of equations for this example system then becomes

$$\frac{\partial \hat{m}_1}{\partial t} = -\gamma\mu_0[\hat{m}_1 \times \bar{H}_{eff}] + \alpha\left[\hat{m}_1 \times \frac{\partial \hat{m}_1}{\partial t}\right] + \frac{\bar{I}_{13\perp}}{eN_s}$$ (37)

$$\frac{\partial \hat{m}_2}{\partial t} = -\gamma\mu_0[\hat{m}_2 \times \bar{H}_{eff}] + \alpha\left[\hat{m}_2 \times \frac{\partial \hat{m}_2}{\partial t}\right] + \frac{\bar{I}_{23\perp}}{eN_s}$$ (38)

where the voltages are derived from equation (36).

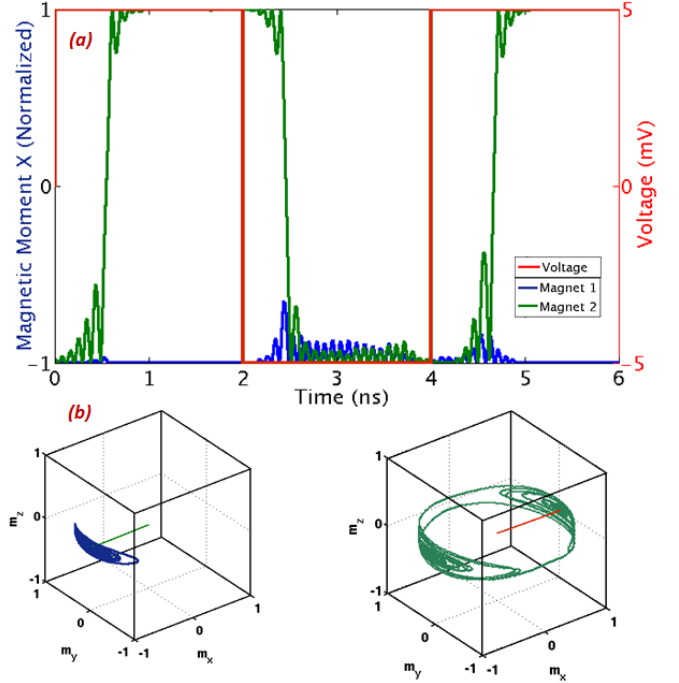

Fig. 10. a) Transient self-consistent simulation of a spin circuit device b) Trajectory of the magnetic moment of the nanomagnets

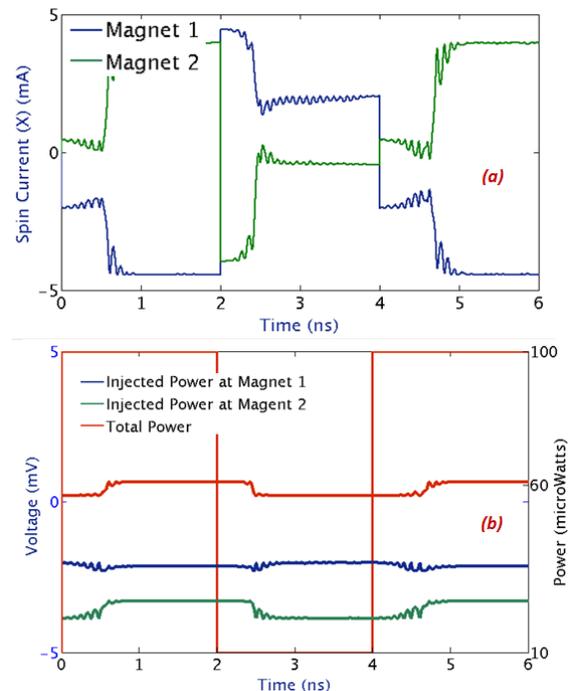

Fig. 11. a) Spin current via input and output magnets. B) The instantaneous power through the spin logic device.

We numerically simulated the spin logic circuit using self-consistent solution of the spin-MNA equation with stochastic LLG equation. The specific parameters used for this example are shown in Table 5 for reproducing the results. We note that the stochastic nature of the LLG equations produces an inherent variability in operation of spin devices.

We show the basic operation of the device in Fig. 10. In Fig. 10 (a) we show the dynamics of the magnetic moment of the nanomagents for various applied voltages. During the time from 0ns to 2ns, a positive voltage is applied at node 3, the device acts as an inverting gate during this interval. The device responds over a time of ~0.5ns and the output flips to a state opposite of the input magnet. At 2ns, the supply voltage is flipped to – 5mV, the device acts as a non-inverting gate in this interval and the output responds at ~2.5ns and flips to a state parallel to the input magnet. The trajectories of the magnets' magnetic moments are shown in Fig. 10 (b).

We now extract the spin and charge currents via the circuit using the branch conductances. The spin currents along the nanomagnet easy axis are show in Fig. 11 a. The total energy dissipation of the device can now be calculated as the total charge current sourced from the supply times the supply voltage. The total electrical power of the device can be extracted as shown in Fig. 11b.

This circuit modeling described here shows that the spin logic circuit, using table 5 parameters, operates with transient energy-delay metrics of ~ 30 fJ/bit with up to a 2 GHz response and with zero leakage power (ignoring the overheads). Compared to a CMOS technology with a 60 nW/transistor idle power [74], non-volatile spin logic with ~0 W state hold power and 7 year retention time will outperform CMOS logic in idle power by several orders of magnitude. We note the leakage power of the MPU units have approached 50% over the past few years [75]. For transient energy-delay improvement, a physically realistic simulation framework, along with improved spin-electrical transduction methods, materials and anisotropy engineering, can address this gap in developing device and circuit topologies to approach the fundamental performance of the spintronic devices and circuits for beyond CMOS era.

## IX. CONCLUSION

In summary, we describe a spin circuit framework combining spin transport with generalized modified node analysis to enable SPICE for spintronic circuit analysis. The framework handles both magnetic and non-magnetic components with the commonly used MNA methodology. This framework will enable study of spintronics for logic, interconnect, memory and hybrid integration of spin devices with CMOS circuits. The ability to synthesize and analyze spintronic CMOS integrated circuits will enable and accelerate the study of spintronics with the potential attributes of non-volatility, superior energy-delay, higher logical efficiency, recofigurability and suitability for novel computational architectures and logic-memory paradigms.

## APPENDIX A: G-MATRIX ELEMENTS OF A NON-MAGNETIC DISTRIBUTED CHANNEL

Spin-dependence conduction can be described in a continuous medium (as opposed to a lumped-element circuit) via the drift-diffusion equations [35]. In the following, we provide the derivation of normal magnet (NM) spin conductance matrix introduced in [18]. We show the spin conductances for currents along one direction (x) and one direction (s) of spin, and then generalize it to arbitrary directions of spins. Then the drift-diffusion equations for the current density J, spin current density $J_s$, voltage V, and spin voltage $V_s$ in a non-magnetic material are

$$J = \sigma \frac{dV}{dx} \quad (A1)$$

$$J_s = \sigma \frac{dV_s}{dx} \quad (A2)$$

Current continuity implies (Kirchhoff current law)

$$\frac{dJ}{dx} = 0 \quad (A3)$$

$$\frac{dJ_s}{dx} = \frac{\sigma}{\lambda^2} V_s \quad (A4)$$

where the conductivity is σ, and the spin diffusion length is $\lambda = \sqrt{D\tau_{sf}}$. A general solution for these equations is J=Constant

$$V_s = a\exp(x/\lambda) + b\exp(-x/\lambda)$$
$$(B5)$$
$$J_s = \frac{\sigma a}{\lambda}\exp(x/\lambda) - \frac{\sigma b}{\lambda}\exp(-x/\lambda) \quad (A5)$$

Where the boundary conditions set the coefficients a, b. The specific solution for the uniform conductor of length L is

$$J = \sigma(V_2 - V_1)/L \quad (A6)$$
$$V_{s1} = a/E + bE \quad (A7)$$
$$V_{s2} = aE + b/E \quad (A8)$$
$$J_{s,in} = \frac{\sigma a}{\lambda E} - \frac{\sigma bE}{\lambda} \quad (A9)$$
$$J_{s,out} = \frac{\sigma aE}{\lambda} - \frac{\sigma b}{\lambda E} \quad (A10)$$

where we designate $E = \exp(L/(2\lambda))$;

From these equations, and for the cross-sectional area $A$ of the conductor

$$G = \sigma A/L \quad (A11)$$

According to the above notation, the following relations must be satisfied for the Π-network

$$J_{s,in} = J_{sp} - J_{s1} \quad (A12)$$
$$J_{s,out} = J_{sp} + J_{s2} \quad (A13)$$
$$AJ_{s1} = G_{sf\pi}V_{s1} \quad (A14)$$
$$AI_{s2} = G_{sf\pi}V_{s2} \quad (A15)$$
$$AJ_{p,s} = G_{se\pi}(V_{s2} - V_{s1}) \quad (A16)$$

for all values of the boundary conditions (specified by a, b). Taking a special case of a=0, b=1, we arrive at the equations

$$\frac{A\sigma E}{\lambda} = G_{se\pi}\left(E - \frac{1}{E}\right) + G_{sf\pi}E \quad (A17)$$

$$\frac{A\sigma}{\lambda E} = G_{se\pi}\left(E - \frac{1}{E}\right) - G_{sf\pi}\frac{1}{E} \quad (A18)$$

which have the solution:

$$G_{se\pi} = \frac{\sigma A}{\lambda}\cosech\left(\frac{L}{\lambda}\right) \quad (A19)$$

$$G_{sf\pi} = \frac{\sigma A}{\lambda}\tanh\left(\frac{L}{2\lambda}\right) \quad (A20)$$

Hence, the conductance for the series branch of the Π network is [18]:

$$G_{se\pi} = \begin{bmatrix} \frac{A_{ch}}{\rho_{ch}L_c} & 0 & 0 & 0 \\ 0 & \frac{A_{ch}}{\rho_{ch}\lambda_s}\csch\left(\frac{L_c}{\lambda_s}\right) & 0 & 0 \\ 0 & 0 & \frac{A_{ch}}{\rho_{ch}\lambda_s}\csch\left(\frac{L_c}{\lambda_s}\right) & 0 \\ 0 & 0 & 0 & \frac{A_{ch}}{\rho_{ch}\lambda_s}\csch\left(\frac{L_c}{\lambda_s}\right) \end{bmatrix} \quad (A21)$$

The conductance for the parallel branch of the Π network is:

$$G_{sf\pi} = \begin{bmatrix} 0 & 0 & 0 & 0 \\ 0 & \frac{A_{ch}}{\rho_{ch}\lambda_s}\tanh\left(\frac{L_c}{2\lambda_s}\right) & 0 & 0 \\ 0 & 0 & \frac{A_{ch}}{\rho_{ch}\lambda_s}\tanh\left(\frac{L_c}{2\lambda_s}\right) & 0 \\ 0 & 0 & 0 & \frac{A_{ch}}{\rho_{ch}\lambda_s}\tanh\left(\frac{L_c}{2\lambda_s}\right) \end{bmatrix} \quad (A22)$$

APPENDIX B: Π TO T EQUIVALENT CIRCUIT TRANSFORMATION

Spin-dependent conductance with spin relaxation can be equivalently represented by Π-shaped (Fig. 4b) or T-shaped (Fig. 4c) networks. Here we derive a general relation between these two approaches.

In both cases the vector voltages at the terminals, V1 and V2 are the same. In the Π-network, the spin-relaxation currents are

$$I_1 = G_{sf\pi}V_1 \quad (B1)$$
$$I_2 = G_{sf\pi}V_2 \quad (B2)$$

& the current passing through the center conductance is

$$I_p = G_{se\pi}(V_2 - V_1) \quad (B3)$$

They are related to the total currents entering and leaving the network as follows

$$I_{in} = I_p - I_1 \quad (B4)$$
$$I_{out} = I_p + I_2 \quad (B5)$$

Using these relations in (C4, C5) is convenient to express the sum and the difference of the in- and out-currents.

$$I_{out} - I_{in} = G_{sf\pi}(V_2 + V_1) \quad (B6)$$
$$I_{out} + I_{in} = (2G_{se\pi} + G_{sf\pi})(V_2 - V_1) \quad (B7)$$

Similarly we obtain for the T-network; the current from the middle node to the ground is related to the vector voltage at this node

$$I_m = G_{sfT}V_m \quad (B8)$$

and the in- and out-currents are

$$I_{in} = G_{seT}(V_m - V_1) \quad (B9)$$
$$I_{out} = G_{seT}(V_2 - V_m) \quad (B10)$$

The current conservation results in

$$I_{out} = I_{in} + I_m \quad (B11)$$

The voltage at the middle node is thus related (for non-zero spin relaxation) to the in- and out-currents

$$V_m = G_{seT}^{-1}(I_{out} - I_{in}) \quad (B12)$$

As before, we express via the unity matrix $I$

$$(I + 2G_{seT}G_{sfT}^{-1})(I_{out} - I_{in}) = G_{seT}(V_2 + V_1) \quad (B13)$$
$$I_{out} + I_{in} = G_{seT}(V_2 - V_1) \quad (B14)$$

Since the two expressions for in- and out-currents should be equivalent, the following relations between non-zero conductances must hold

$$G_{seT} = 2G_{se\pi} + G_{sf\pi} \quad (B15)$$
$$G_{sf\pi}^{-1} = G_{seT}^{-1} + 2G_{sfT}^{-1} \quad (B16)$$

Simple algebraic manipulations permit the expression of the T-conductances in terms on π-conductances

$$G_{sfT} = G_{sf\pi}G_{se\pi}^{-1}G_{sf\pi} + 2G_{sf\pi} \quad (B17)$$

Note that this derivation is not trivially reduced to the results of the traditional electronic network theory, because the conductances are matrices rather than scalars.

APPENDIX C: G-MATRIX ELEMENTS OF FM EXPRESSED AS SPIN REFLECTION AND TRANSMISSION COEFFICIENTS

*A. 4X4 conduction matrix elements of a ferromagnetic metal in contact with a normal metal expressed in terms of spin reflection and transmission coefficients*

Earlier, we described the conduction matrix elements of a FM as phenomenological constants extracted from experimental properties of the FM-NM conduction. Here, we provide a description of the conduction matrix elements of a FM that is derived from an ab initio approach [18, 38]. The conduction matrix

$$G_{FM} = \begin{bmatrix} G & \alpha G & 0 & 0 \\ \alpha G & G & 0 & 0 \\ 0 & 0 & G_{SL} & G_{FL} \\ 0 & 0 & -G_{FL} & G_{SL} \end{bmatrix} \quad (C1)$$

can also be written in terms of spin scattering conductance elements as

$$G_{FM} = \begin{bmatrix} G_{\uparrow\uparrow} + G_{\downarrow\downarrow} & G_{\uparrow\uparrow} - G_{\downarrow\downarrow} & 0 & 0 \\ G_{\uparrow\uparrow} - G_{\downarrow\downarrow} & G_{\uparrow\uparrow} + G_{\downarrow\downarrow} & 0 & 0 \\ 0 & 0 & 2\operatorname{Re}G_{\uparrow\downarrow} & 2\operatorname{Im}G_{\uparrow\downarrow} \\ 0 & 0 & -2\operatorname{Im}G_{\uparrow\downarrow} & 2\operatorname{Re}G_{\uparrow\downarrow} \end{bmatrix} \quad (C2)$$

where $G_{\uparrow\uparrow}$, $G_{\downarrow\downarrow}$ and $G_{\uparrow\downarrow}$ are the matrix elements derived from spin scattering at the FM-NM interface. The conduction matrix elements of an FM-NM interface can be described in terms of the reflection and transmission properties of the spin up and spin down electrons incident from an NM to FM [38].

$$G^{\uparrow\uparrow} = \frac{e^2}{h} \sum_{n \in NM} \sum_{m \in FM} |t_\uparrow^{nm}|^2 \quad (C3)$$

$$G^{\downarrow\downarrow} = \frac{e^2}{h} \sum_{n \in NM} \sum_{m \in FM} |t_\downarrow^{nm}|^2 \quad (C4)$$

$$G^{\uparrow\downarrow} = G^{\downarrow\uparrow*} = \frac{e^2}{h} \sum_{n \in NM} \left(1 - \sum_{m \in NM} r_\uparrow^{nm} r_\downarrow^{nm*}\right) \quad (C5)$$

where $e^2/h$ is the conductance per spin of a ballistic channel with ideal contacts [38]; $t_\downarrow^{nm}, t_\uparrow^{nm}$ are the transmission coefficients for up and down spin electrons from NM to FM; $r_\uparrow^{nm}, r_\downarrow^{nm}$ are the reflection coefficients of the up and down spin electrons at the FM-NM interface; $n$ is the number of modes in the NM, $m$ is the number of modes in the FM. The number of modes in a metal *NM* can in-turn be written from the metal's Fermi wave vector $k_f$ [48].

It has been argued that $r_\uparrow^{nm}, r_\downarrow^{nm}$ are close to zero for many material systems [38], which simplifies the spin torque conductance to $G_{SL}$ to

$$G_{SL} = 2\,\text{Re}\,G^{\uparrow\downarrow} = \frac{2e^2}{h} n = \frac{2e^2}{h} \frac{A k_f^2}{4\pi} \quad (C6)$$

### APPENDIX D: G-MATRIX OF FREE FM LAYER

*A. Derivation for free magnetic layer conduction matrix*

Here we derive the G-Matrix for a free layer FM with an arbitrary magnetic moment direction ($\hat{m}$) (see Fig. 7). Let φ be the angle of the magnetic moment with z-axis and θ be the angle of the projection of ($\hat{m}$), with x-axis. We can write $\hat{m}$ in co-ordinate system xyz as:

$$\hat{m} = \cos\theta \sin\varphi.\hat{x} + \sin\theta \sin\varphi.\hat{y} + \cos\varphi.\hat{z} \quad (D1)$$

Let us choose a new co-ordinate system XYZ such that, the $\hat{m}$ is collinear with the new X-axis. (Refer to section V)

$$\hat{X} = \hat{m} \quad (D2)$$

$$\hat{Y} = -\left(\hat{X} \times \hat{x}\right) / \|\hat{X} \times \hat{x}\| \quad (d3)$$

$$\hat{Z} = \hat{X} \times \hat{Y} \quad (D4)$$

In the new co-ordinate system:

$$\vec{I}_{XYZ} = G_{FN}(\hat{x}) \vec{V}_{XYZ} \quad (D5)$$

where $G_0$ is the matrix described in section V. Let us substitute

$$\vec{I}_{XYZ} = R \vec{I}_{xyz} \quad (D6)$$

$$\vec{V}_{XYZ} = R \vec{V}_{xyz} \quad (D7)$$

And rearrange to obtain the current, voltage relation in the xyz co-ordinate system. We obtain

$$\vec{I}_{xyz} = R^{-1} G_{FN}(\hat{x}) R \vec{V}_{xyz} \quad (D8)$$

Hence, the conductance matrix for an FM with magnetic moment along an arbitrary direction is given by:

$$G_{FN}(\hat{m}) = R^{-1}(\hat{m}) G_{FN}(\hat{x}) R(\hat{m}) \quad (D9)$$

### APPENDIX E: CONVERSION FROM SPINOR BASIS TO VECTOR BASIS

*A. Conversion between spinor spin current/voltage basis to 4 component vector current/voltage basis.*

The derivations for magneto-electronic circuit theory are often performed in the spinor basis for the electrons [39]. For convenience we list the conversion from spinor basis to Cartesian vector basis as well as 4-component current basis. The current in a spinor basis can be written as [39],

$$\hat{I} = (I_c \hat{1} + \vec{I}_s.\vec{\sigma})/2 \quad (E1)$$

where $\vec{\sigma}$ is the Pauli spin matrix :

$$\vec{\sigma} = \sigma_x \hat{x} + \sigma_y \hat{y} + \sigma_z \hat{z} \quad (E2)$$

Which yields,

$$\hat{I} = \begin{bmatrix} I_{\uparrow\uparrow} & I_{\uparrow\downarrow} \\ I_{\downarrow\uparrow} & I_{\downarrow\downarrow} \end{bmatrix} = \begin{bmatrix} I_c + I_z & I_x - iI_y \\ I_x + iI_y & I_c - I_z \end{bmatrix} \quad (E3)$$

Hence, the 4-component current vector can be derived from the spinor current as follows:

$$\bar{I} = \frac{1}{2} \begin{bmatrix} I_{\uparrow\uparrow} + I_{\downarrow\downarrow} \\ I_{\uparrow\downarrow} + I_{\downarrow\uparrow} \\ i(I_{\uparrow\downarrow} - I_{\downarrow\uparrow}) \\ I_{\uparrow\uparrow} - I_{\downarrow\downarrow} \end{bmatrix} \quad (E4)$$

Similarly, 4-component vector spin voltage can be derived from spinor voltage as follows:

$$\bar{V} = \frac{1}{2} \begin{bmatrix} V_{\uparrow\uparrow} + V_{\downarrow\downarrow} \\ V_{\uparrow\downarrow} + V_{\downarrow\uparrow} \\ i(V_{\uparrow\downarrow} - V_{\downarrow\uparrow}) \\ V_{\uparrow\uparrow} - V_{\downarrow\downarrow} \end{bmatrix} \quad (E5)$$

### APPENDIX F: STOCHASTIC LLG EQUATIONS

*A. Thermal noise of nanomagnets*

The dynamics of nanomagnets are strongly affected by the thermal noise. Thermal noise in a nanomagnet manifests as fluctuations to the internal anisotropic field [51-54]. The thermal noise can be considered as a result of the microscopic degrees of freedom of the conduction electrons and the lattice of the ferromagnetic element [51].

At room temperature T, the thermal noise is described by a Gaussian white noise (with a time domain Dirac-delta auto-correlation). The noise field acts isotropically on the magnet. In presence of the noise, the LLG equation can be written as

$$\frac{\partial \hat{m}}{\partial t} = -\gamma \mu_0 [\hat{m} \times \bar{H}_{eff}(T)] + \alpha \left[\hat{m} \times \frac{\partial m}{\partial t}\right] - \frac{\vec{I}_\perp}{eN_s} \quad (F1)$$

Where we modified (34) by adding temperature dependence. The internal field is described as:

$$\bar{H}_{eff}(T) = \bar{H}_{eff} + (H_i \hat{x} + H_j \hat{y} + H_k \hat{z}) \quad (F2)$$

$$\langle H_l(t) \rangle = 0 \quad (F3)$$

$$\langle H_l(t)H_k(t')\rangle = \frac{2\alpha k_B T}{\mu_0^2 \gamma M_s V}\delta(t-t')\delta_{lk} \quad \text{(F4)}$$

The initial conditions of the magnets should also be randomized to be consistent with the distribution of initial angles of magnet moments in a large collection of magnets. At temperature T, the initial angle of the magnets follows [52]:

$$\langle \theta^2 \rangle = \frac{kT}{M_s V \mu_0 H_{ani}} \quad \text{(F5)}$$

*B. Numerical methods for stochastic LLG equations*

An accurate choice of the method for integration of the stochastic LLG equation is essential since a) the stochastic differential equations (SDE) require careful handling of the order of integration [67, 68] 2) a multiplicative white noise requires an appropriate choice of calculus [53, 69]. The appropriate model for direct integration of SDE are usually first order integration methods such as Euler & Heun. Even though higher order methods have been proposed in the literature, the accuracy and applicability for realistic SDEs have been questioned [67, 68]. Hence, a first order integration method with a fixed time step is generally recommended [53].

Secondly, Stratonovich calculus is used for interpreting the multiplicative white noise. We used a mid-point integration method [53] to apply the Stratonovich calculus while integrating the LLG equation. The discretized integration rule is

$$y_{n+1} = y_n + \Delta t . f(t_n + \Delta t/2, y_n + \Delta t/2 . f(t_n, y_n)) \quad \text{(F6)}$$

where $y = d\hat{m}/dt$. The variance of the noise varies depending on the time step size. The discretization was performed internally using a matlab implicit self-consistent solver.

APPENDIX G: EQUIVALENT CONDUCTANCES SERIES AND PARALLEL

The equivalent conductance of a two spin conductance elements connected in parallel is

$$G_\| = G_1 + G_2 \quad \text{(G1)}$$

$$G_{se} = (G_1 + G_2)^{-1} G_1 G_2 \quad \text{(G2)}$$

The rule for voltage division says, voltage across conductance $i=1, 2$ is

$$\Delta V_i = (G_1 + G_2)^{-1} G_i V \quad \text{(G3)}$$

APPENDIX H: WRITING SPIN MNA EQUATIONS, NOTATIONS

Rules for writing the spin MNA equations are very similar to the rules for regular MNA [63, 65]. However, care should be taken to include the spin current conservation law described in section III. Also, 4X4 identity matrices are to be used where appropriate to indicate the voltage sources,

To apply the spin-MNA method to a circuit with n nodes with m voltage sources, the steps are

1. *Number the nodes:* Select a reference node (usually the ground, numbered 0) and name the remaining n-1 nodes. Also label currents through each current source.
2. *Name the currents:* Assign a name to the current through each voltage source flowing from positive node to negative node of the source.
3. *Apply spin KCL:* Apply Spin current conservation law at each node with current into the node to be positive.
4. Write an equation for spin voltage at each spin voltage source.
5. Rearrange the equations to the form of equation (35).
6. Invert the **A** matrix and multiply with **Z** to obtain the voltages of the nodes and current of the sources.
7. Using conductance matrix to calculate the relevant currents.

APPENDIX I: COMMENT ON EXTENSION FOR SEMICONDUCTOR SPIN TRANSPORT

We briefly comment on the extendibility of the present approach to semiconductor spin transport [70]. The major modifications required to handle semiconductors using conduction matrix approach are as follows a) conductance matrix for describing drift-diffusion spin current [71] b) description of dilute magnetic semiconductors c) appropriate definition of spin currents in presence of spin orbit coupling [72, 73].

ACKNOWLEDGMENT

We acknowledge the discussions with Prof. Supriyo Datta, Srikant Srinivasan, and Behtash Behin-Aein, at Purdue University and Prof. Sayeef Salahuddin at UC Berkeley.

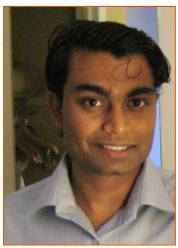

**Sasikanth Manipatruni** (M'07) is a research scientist in the Exploratory Integrated Devices and Circuits group in Intel Components Research, Technology and Manufacturing Group at Intel Corporation. He is working on emerging novel devices to identify technology options for beyond CMOS logic technologies. Dr. Manipatruni joined Intel in 2011. Before that he was at General Electric (GE) Global Research Center working in nanophotonic interconnects for massively parallel magnetic resonance imaging (MRI) and demonstrated the first analog-RF optical MRI system at 3T magnetic fields.

Born in Vizianagaram, India, he obtained his Ph.D. from Cornell University working with Prof. Michal Lipson in silicon photonics and opto-mechanics. During his PhD, he was co-inventor of several silicon photonic devices including the first 18 Gb/s micro-ring modulator, first GHz poly-silicon modulator, hitless broadband EO switches & integrated silicon nanophotonic link. He graduated from Indian Institute of Technology (IIT) Delhi at the top of his class in EE in 2005 with the best hardware thesis project award. He was a KVPY national science fellow of the Indian Institute of Sciences (IISc) (1999-2001); worked at Swiss Federal Institute of Technology (ETH), Zurich in 2004; and Inter University center for astronomy and astrophysics (IUCAA) in 2001. He has more than 10 patent applications in nano-photonics and MRI, 40 peer reviewed journal and conference papers. He serves as a peer reviewer for OSA, IEEE and Nature Photonics. His business plan for fab-less photonics was awarded the first place in Asia-Pacific at Asia Moot Corp 2008 and selected as an outstanding product at World Moot Corp 2009.

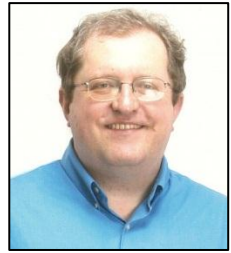

**Dmtri E. Nikonov** ((M'99-SM'06) received M.S. in Aeromechanical Engineering from the Moscow Institute of Physics and Technology in 1992 and Ph.D. in Physics from Texas A&M University in 1996, where he participated in the demonstration of the world's first laser without population inversion.

He joined Intel Corporation in 1998 and is presently a Research Scientist in the Components Research group in Hillsboro, Origin. He is responsible for simulation and benchmarking of beyond-CMOS logic devices and for managing joint research programs with universities on nanotechnology and exploratory devices. From 1997 to 1998 he was a research engineer and lecturer at the Department of Electrical and Computer Engineering of University of California Santa Barbara. In 2006 he was appointed Adjunct Associate Professor of Electrical and Computer Engineering at Purdue University. Dmitri was a finalist of the Best Doctoral Thesis competition of the American Physical Society in 1997. He has 60 publications in refereed journals in quantum optics, lasers, nanoelectronics and spintronics, and 35 issued patents in optoelectronics, integrated optic and spintronic devices.

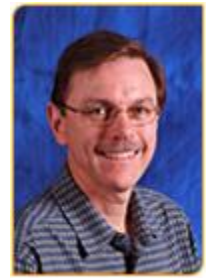

**Ian. A. Young** (M'78–SM'96–F'99) is a Senior Fellow and Director of Exploratory Integrated Devices and Circuits group in Intel Components Research, Technology and Manufacturing Group at Intel Corporation. He is responsible for defining and developing future circuit directions with emerging novel devices and identifying leading options for devices and interconnects to manufacture solid-state integrated circuits in the post-CMOS era. Dr. Young joined Intel in 1983. Starting with the development of circuits for a 1 Mb DRAM, and the world's first 1 μm 64 K SRAM, he then led the design of three generations of SRAM products and manufacturing test vehicles, and developed the original Phase Locked Loop (PLL) based clocking circuit in a microprocessor while working on the 50 MHz Intel 486™ processor design. He subsequently developed the core PLL clocking circuit building blocks used in each generation of Intel microprocessors through the 0.13 μm 3.2 GHz Pentium 4. Dr. Young has developed a number of optimization metrics for process technology development, including the transistor performance metric that provided a link between processor performance and basic transistor parameters, as well as back-end metal interconnect architecture.

Born in Melbourne, Australia, he received his bachelor's and master's degrees in electrical engineering from the University of Melbourne, Australia, in 1972 and 1975. He received his Ph.D. in electrical engineering from the University of California, Berkeley in 1978. Prior to Intel, Dr. Young worked on analog/digital integrated circuits for Telecommunications products at Mostek Corporation (United Technologies), as well as an independent design consultant.

Dr. Young was a member of the Symposium on VLSI Circuits Technical Program Committee from 1991 to 1996, serving as the Program Committee Chairman in 1995/1996, and the Symposium Chairman in 1997/1998. He was a member of the ISSCC Technical Program Committee from 1992 to 2005, serving as the Digital Subcommittee Chair from 1997 to 2003, the Technical Program Committee Vice-chair in 2004 and Chair in 2005. He was Guest Editor for the April 1997, April 1996 and December 1994 issues of the JSSC. He has served as an elected member of the SSCS Adcom from 2006 to 2009. Young received the International Solid-State Circuits Conference's 2009 Jack Raper Award for Outstanding Technology-Directions, for the paper entitled "Optical I/O Technology in Tera-Scale Computing". Dr. Young is a Fellow of the IEEE. He holds 45 patents in integrated circuits and has authored or co-authored over 40 technical papers.